\documentclass{aastex62}

\usepackage[utf8x]{inputenc}
\usepackage[english]{babel}
\usepackage{fontenc}
\usepackage{graphicx}
\graphicspath{{./}}

\usepackage{natbib}
\usepackage{amsmath}
\usepackage{amsfonts}
\usepackage{amssymb}
\usepackage{mathrsfs}

\usepackage{wasysym}  

\usepackage{siunitx}

\begin{document}

\title{HydroSyMBA: a 1D hydrocode coupled with an N-body symplectic integrator.}

\author{J.~Salmon}
\affil{Southwest Research Institute\\Planetary Science Directorate\\1050 Walnut Street, Suite 300, Boulder, CO 80302, USA}
\email{julien@boulder.swri.edu}
\author{R.~M.~Canup}
\affil{Southwest Research Institute\\Planetary Science Directorate\\1050 Walnut Street, Suite 300, Boulder, CO 80302, USA}
\email{robin@boulder.swri.edu}

\begin{abstract}
The numerical modeling of co-existing circumplanetary disks/rings and satellites is particularly challenging because each part of the system requires a very different approach. Disks are generally well represented by a fluid-like dense medium, whose evolution can be calculated by a hydrocode. On the other hand, the orbital evolution of satellites is generally performed using \textit{N}-body integrators. We have developed a new numerical model that combines a 1-dimensional hydrocode with the \textit{N}-body integrator SyMBA. The disk evolves due to its viscosity, and resonant torques from satellites. The latter is applied to the satellites as an additional ``kick'' to their accelerations. The integrator also includes the ability to spawn new moonlets at the disk's outer edge if the latter expands beyond a material-dependent Roche limit, as well as the effects of tidal dissipation in the planet and/or the satellite on the satellite orbits. The resulting integrator allows one to accurately model the evolution of an inner circumplanetary disk, and the formation of satellites by accumulation of disk material, all within a single self-consistent framework. Potential applications include the formation of Earth's Moon, the evolution of the inner Saturn system, the martian and uranian moons, and compact exoplanetary systems.
\end{abstract}


\section{Introduction}
The formation and the evolution of planetary satellites is a fundamental topic in planetary science. The history of these systems is closely tied to that of the planet they orbit, such that understanding satellites is an indirect way of studying planet formation in general, and the history of our Solar System in particular. The irregular satellites, such as Triton around Neptune, are likely objects that formed in the outer Solar System, were later scattered inward, passed close to the planet and got captured via gravitational interactions. The \textit{regular} satellites likely formed from a disk around the planet itself.

Planets form inside the proto-planetary disk, a giant disk of gas and dust orbiting the Sun. If the planet grows massive enough, some gas and dust from the main nebula can become bound to the planet and orbit it, forming a \textit{subnebula}. We believe that the Galilean Satellites formed by accretion of gas and dust in Jupiter's subnebula \citep[e.g.][]{canup06}. The Earth's Moon likely formed by accretion from a disk, itself the product of a giant impact suffered by Earth toward the end of its formation \citep{cameron76,salmon12}. Some of Saturn's satellites may be the \textit{children} of Saturn's rings, as they may have assembled by accumulation of material from rings that were much more massive in the past \citep{canup10,charnoz10,charnoz11,salmon17}. Phobos and Deimos, satellites of Mars, are often viewed as captured asteroids, but their equatorial, quasi-circular and low-inclination orbits point to formation from an ancient circum-Mars ring \citep{rosenblatt16,Hesselbrock17,hyodo17,canup18}. Uranus and Neptune's regular satellites are also compatible with formation from a ring \citep{crida12}.

\begin{figure}[h!]
	\begin{center}
	\includegraphics[width=17cm]{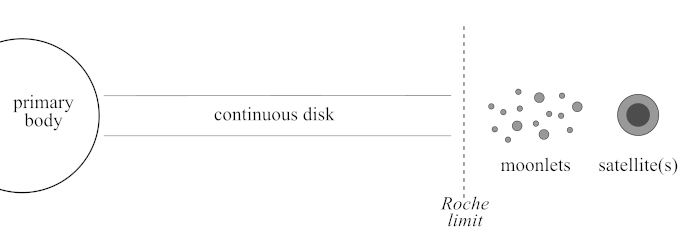}
	\caption{Schematic of the systems we model in our numerical code. The primary (e.g. a planet) is surrounded by a continuous disk that can be in a condensed state, or vaporized, or a mixture of melt and vapor. Beyond the Roche limit, the disk fragments into discrete objects that can accumulate into satellites through collisions. \label{figSchematic}} \renewcommand{\baselinestretch}{0.9}
	\end{center}
\end{figure}

Of particular importance for these systems is the dynamical transition that separates an inner disk, or ring, from the region of growing satellites: the Roche limit. This distance is a function of the densities of the central body and the accreting disk material. Inside this distance, the differential gravitational force from the primary body across two colliding bodies is strong and generally prevents the bodies from ``sticking'' to each other by mutual gravity. As a result, only small objects, which are held together by the strength of the material they are made from rather than from their gravity, are generally present. Outside of the Roche limit, two colliding bodies can remain bound by their mutual gravity, allowing larger objects to grow, eventually forming satellites (Figure \ref{figSchematic}). A prime example is Saturn's system. Saturn's rings are composed primarily of water ice, setting the Roche limit at $\sim 140000$ km. Inside this distance lie the rings, composed of particles of order $\SI{1}{\meter}$. In the transition region around this distance lies the F ring, which has a spiral structure and within which the appearance and destruction of clumps of material are observed. Beyond this distance we find the many Saturnian satellites.

While circumplanetary disks and satellites appear to share a common history, these systems are inherently difficult to study in a single, self-consistent model, given the different numerical approaches necessary to correctly model the rings and the satellites. Modeling the entirety of a cold particulate disk, such as Saturn's rings, by a collection of individual particles would require one to include billions of small objects, which would rapidly limit computational efficiency. However, if the disk is dense enough, then the evolution of the disk is dominated by collective effects, so that it becomes possible to model it as a continuous medium \citep{spahn06}, which can be represented on a discrete grid with a few $10^2$ to $10^4$ radial or annular cells , greatly improving numerical efficiency. Hydrocodes are generally the tool of choice to study the evolution of circumplanetary disks.

To model the growth of individual satellites, it is necessary to accurately integrate their orbital evolution, which is dominated by few-body gravitational interactions, with the planet and possibly other satellites. This is generally done using \textit{N}-body integrators, which model each satellite individually and evolve its orbit due to the gravitational potential of the planet and that of other orbiting bodies during close encounters. Most state-of-the-art integrators can efficiently integrate up to 10000 bodies over $\sim10^7$ orbits.

Due to the large difference in modeling needs, most numerical integrators use simplified approaches to model part of the system, which generally prohibits including important physical processes. For instance, the integrator used in \cite{charnoz10} did not explicitly model the interactions between growing satellites, such that important interactions such as capture in mean-motion resonances could not effectively occur. On the other hand, the integrator used in \cite{salmon12} did not resolve the radial structure of the disk, such that the resonant torque from an outer moonlet was not deposited at the actual position of the resonance, but instead was applied at the disk's outer edge, which would strongly affect the outward spreading of the disk in that region.

In an attempt to remedy the limitations of current integrators, we have developed a new code that combines a hydrocode component to model a continuous inner disk, and an \textit{N}-body integrator to fully track the orbital evolution of discrete outer bodies. While \textit{Fargo-SyMBA} \citep{morbidelli12a} uses a similar approach, its numerical complexity and the absence of treatment of accretion processes at the Roche limit make it unsuitable for studying the formation of satellites from a spreading, Roche-interior circumplanetary disk. The disk section in our new model is broadly inspired from that in \cite{charnoz10}, although we have largely rewritten the disk-satellite interaction portion to address issues with conservation of the system's angular momentum. The N-body part is handled by a modified version of the symplectic integrator SyMBA \citep{duncan98}, in which the particle-particle interactions remain unchanged, but we have added additional accelerations on the bodies due to interactions with the disk and tidal dissipation. 

The resulting code allows one to model the full system of rings/disk and satellites with a tailored approach for each part, and to include the most important dynamical processes. While in the following we focus our presentation on systems where the primary is a planet, the latter could also, for example, be a star in systems with very close-in orbiting exoplanets \citep[e.g.][]{vanlieshout18}, many of which have been identified in the last decade, in particular via the Kepler mission. In Section 2 we describe the details of the integrator. In Section 3 we present some test results, and in Section 4 we discuss some potential future developments.

\section{Relevant physical processes}
\subsection{Viscous spreading of a disk}
The numerical study of a massive disk of particles, such as Saturn's rings, is a technical challenge. High-resolution modeling, in which one considers the individual particles composing the disk, is strongly limited by numerical capabilities. As a result, studies limit themselves to a small \textit{patch} of the rings and perform integration over a few orbits \citep[e.g.][]{salo92,salo95,wisdom88}. These studies are fundamental, as they provide a better understanding of the basic physics affecting the evolution of the rings.
 
One of the most basic processes driving the evolution of a disk of material is the redistribution of angular momentum via collisions. Particles on close, nearly circular orbits move at slightly different velocities: inner particles move faster and can then overtake and collide with outer particles. The angular momentum of the pair of particles is redistributed such that the outer (inner) particle gains (loses) angular momentum. Since angular momentum is proportional to the square root of the particles' distance to the planet, the outer (inner) particle moves outward (inward). As a result, the disk spreads \citep[e.g.][]{lynden74}.
 
Collisions (or any friction that dissipates energy) thus cause a global transfer of angular momentum from the inner to the outer regions of the disk. By analogy with the transfer of angular momentum in a non-ideal fluid, the efficiency of the transfer is modeled by a \textit{viscosity}, whose value is generally a function of distance and local mass of the disk. This allows one to model the disk not by a collection of discrete objects, but as a continuous medium or \textit{fluid} in which angular momentum is redistributed at a rate specified by an appropriate viscosity. The latter can be modeled by performing local simulations such as those mentioned above, or through analytical treatments.

 \subsection{Accretion outside the Roche limit}
 When two objects collide within the Roche limit, they typically cannot remain bound solely due to gravity as the differential gravitational pull from the primary body is stronger than the gravity between the two objects \citep{canup95}. This prevents large objects from growing, and only small particles can exist, as their integrity is preserved by the strength of the material they are made of. As the disk spreads due to its viscosity, it will bring material through the Roche limit. This material can then coalesce by gravitational instabilities which can cause the particles from the disk to clump into larger objects \citep{goldreich73}. Being outside the Roche limit, these clumps will not be sheared apart by the planet and can form new moonlets that separate from the disk. As these moonlets evolve inside the ``satellite region'' they may collide with each other and progressively grow into larger objects through two-body collisions, potentially forming larger satellites. This growth mechanism has been observed both in numerical \citep{charnoz10} and analytical studies \citep[the ``pyramidal regime'' of][]{crida12}
 
 \subsection{Disk-satellite interactions}
 A third physical process that needs to be considered is the gravitational interactions between the disk and the orbiting moons. For an inner disk, these contribute to transferring angular momentum from the disk to the moons. This results in a contraction of the disk, and an expansion of the satellites' orbits. This occurs through a type of resonance called \textit{Lindblad} resonances, which occur when the ratio of the orbital frequency $\Omega(r)$ at distance $r$ in the disk is a rational multiple of that of the satellite $\Omega_s$: $p\Omega(r) = q\Omega_s$, where $p$ and $q$ are integers greater than 1. First order Lindblad resonances, which occur when $q=p+1$, are the strongest, such that limiting the computation to these generally gives a good approximation to the system's behavior.
 
 \subsection{Tides}
 A fourth process that affects the satellites is tidal dissipation. A satellite will cause a deformation of the planet it orbits via the gravitational attraction it exerts. Since the planet is not a perfect fluid, the deformation will not occur instantaneously in response to the satellite's gravity. As a result, the satellite is not directly above the tidal bulge it generates on the planet. The deformation of the planet changes its gravitational potential, and the lag between the satellite and the position of the bulge creates an asymmetry which gives rise to a torque on the planet, and an equal and opposite torque on the satellite. If the satellite's orbit is such that its angular orbital velocity is faster than the planet's angular rotation velocity, then the satellite leads the bulge and it receives a negative torque which causes its orbit to contract. The planet receives an equal and opposite torque that accelerates its spin. If the satellite orbits more slowly that the planet spins, then it receives a positive torque and its orbit expands, while the planet's spin is decelerated.
 
 Additionally, the planet may also raise tides on the satellite (``satellite tides''). These tend to decrease the satellite's orbital eccentricity and drive its spin state to synchronous rotation (see Section \ref{subsectides}).
 
 \section{Numerical implementation.} 
 \subsection{Basic architecture of the integrator}
 The new integrator is composed of two components: a disk component whose evolution is computed using a hydrocode, and a satellite component using an \textit{N}-body symplectic integrator. The disk is discretized on a 1-dimensional grid (in the radial direction) comprised of N equal-size bins, and at each timestep its surface density is evolved by computing: 1) the mass fluxes between each cell of the grid due to the viscous torque, and 2) the variation of the disk's angular momentum due to resonant torques (if satellites are present). The \textit{N}-body integrator is a modified version of the symplectic integrator SyMBA \citep{duncan88,salmon12,salmon17}. At each timestep it evolves the orbits of discrete objects due to tides, close encounters with other bodies, and resonant torques from the disk.
 
 We have designed the integrator such that SyMBA handles the initialization of the system and serves as the main code to drive the model. At each timestep, the integrator calculates the evolution of the disk for half a timestep $dt$. The latter is performed with a second-order Runge-kutta integrator which evolves the surface density of the disk due to viscous spreading (using Eq. \ref{equ_viscous_spreading}) and resonant interactions with satellites (using Eq. \ref{equ_sat_torque1} and Eq. \ref{equ_sat_torque2}).
 During this operation, it calculates accelerations on each body due to resonant torques. If tides are included, accelerations on each particle due to this process are calculated as well. These accelerations are used to apply an additional ``kick'' to the momentum of the particles, before the core N-body integration of SyMBA is performed. We repeat the same operation for the latter half of the timestep. Thus the additional accelerations are added symmetrically around the main SyMBA operator, so that the disk and tidal accelerations are treated as small ``kicks'' to each object's motion \citep{mcneil05,canup06}. One can express our algorithm using the following operator form:
 
 \begin{equation}
 E_{\rm Disk}\left(\frac{dt}{2}\right) E_{\rm Tides}\left(\frac{dt}{2}\right) E_{\rm SyMBA}\left(dt\right) E_{\rm Tides}\left(\frac{dt}{2}\right) E_{\rm Disk}\left(\frac{dt}{2}\right).
 \end{equation}
 
 \subsection{Viscous spreading}
 The equations of mass and angular momentum conservation read \citep[e.g.][]{pringle81}
 
 \begin{align}
 &r\frac{\partial \sigma}{\partial t}+\frac{\partial\left(r\sigma v_r \right)}{\partial r} = 0\label{eqmasscons}\\
 &\frac{\partial \left(\sigma r^2\Omega\right)}{\partial t}+\frac{1}{r}\frac{\partial\left(\sigma r^3 \Omega v_r \right)}{\partial r} = \frac{1}{r}\frac{\partial}{\partial r}\left(\nu \sigma r^3 \frac{\partial\Omega}{\partial r}\right)\label{eqangmomcons}
 \end{align}
 where $\sigma$, $\nu$ and $v_r$ are the disk's surface density, viscosity and radial velocity, respectively. These quantities are functions of time $t$ and distance $r$. Combining equations \ref{eqmasscons} and \ref{eqangmomcons}, one gets the equation governing the radial evolution of the disk's surface density due to viscous spreading:
 
 \begin{equation}\label{eqsurfdensevol}
\frac{\partial\sigma}{\partial t} = \frac{3}{r}\frac{\partial}{\partial r}\left(\sqrt{r}\frac{\partial\left(\nu\sigma\sqrt{r}\right)}{\partial r}\right)
 \end{equation}
 
 The disk's surface density is evolved using a second-order Runge-Kutta integrator to solve equation \ref{eqsurfdensevol}, using an approach similar to \cite{salmon10}. At each timestep, the surface density in each bin $2\le j\le (N-1)$, centered at distance $R_j$, is evolved by combining the mass fluxes through each edge of the bin:
 \begin{equation}
 \sigma_j(t+dt) = \sigma_j(t)+\frac{F_{j+1}-F_j}{2\pi R_j\Delta R}dt.\label{equ_viscous_spreading}
 \end{equation}
 where the mass flux $F_j$ through the edge $E_j=R_j-\Delta R/2$ of bin j is given by
 \begin{equation}
 F_j=6\pi\sqrt{E_j} \frac{\left(\nu_j\sigma_j\sqrt{R_j}-\nu_{j-1}\sigma_{j-1}\sqrt{R_{j-1}}\right)}{\Delta R}.
 \end{equation}
We then apply user-selected boundary conditions to set $F_1$ and $F_{N+1}$, e.g. ``free-flow'' $(\bf F_1 = F_2,F_{N+1}=F_N)$, ``stop-flow'' $(\bf F_1=0, F_{N+1}=0)$, or ``no-inflow'' $(\bf F_1=0$ if $\bf F_2>0, F_{N+1}=0$ if $\bf F_N<0)$, which allows one to evolve the surface density in the innermost and outermost bins. It is possible to select different boundary conditions for the innermost and outermost bin. We have included various viscosity models in the integrator: viscosity from gravitational instabilities \citep{ward78}, multi-component viscosity for particulate disks \citep{daisaka01}, two-phase protolunar disk \citep{thompson88}, and $\alpha-$prescription \citep{shakura73}.
 
\subsection{Disk-satellite interactions}
Resonant interactions transfer angular momentum between the disk and orbiting bodies. The torque exerted by a satellite on a disk at an (m:m-1) inner Lindblad resonance is given by \citep{goldreich80}:

\begin{equation}
\Gamma_m=-\frac{\pi^2\sigma}{3\omega\Omega_{ps}}\left[r\frac{d\Phi_m}{dr}+\frac{2\Omega}{\omega-\Omega_{ps}}\Phi_m\right]^2,
\end{equation}
 where $\sigma$ is the disk's surface density, $\omega$ is the orbital frequency at distance $r$, $\Omega_{ps}$ is the patern speed of the resonance and $\Phi_m$ is the mth-order Fourier component of the satellite's gravitational potential. In our integrator we limit ourselves to first-order resonances, which are the strongest, and for which $\Omega_{ps}$ is equal to the satellite's orbital frequency $\Omega_S$. The satellite's potential can then be written \citep{goldreich78}:
 
 \begin{equation}
 \Phi_m = -\frac{GM_S}{a_s}b_{1/2}^{(m)}(\alpha),
  \end{equation}
 where $M_S$ and $a_S$ are the satellite's mass and semi-major axis, $\alpha = r/a_S=\left(1-1/m\right)^{2/3}$, and $b_{1/2}^{(m)}(\alpha)$ is the Laplace coefficient of order $1/2$ defined by:
 
 \begin{equation}
 b_i^{(m)}(\alpha)=\frac{2}{\pi}\int_0^\pi\frac{\cos(m\theta)d\theta}{\left(1-2\alpha \cos\theta+\alpha^2\right)^i}
 \end{equation}
 
 At a first-order resonance location we have $\Omega = m\Omega_S/(m-1)$ and the resonant torque can be written
 
 \begin{equation}
 \Gamma_m = \frac{\pi^2}{3}\frac{M_S^2}{M_P}G\sigma a_sc_m,
 \end{equation}
 where $M_P$ is the mass of the central planet, $G$ is the gravitational constant and $c_m = \alpha^{3/2}\left(\alpha\frac{db_{1/2}^{(m)}}{d\alpha} + 2mb_{1/2}^{(m)}\right)^2$. 
 
 We then use the following approximation \citep{goldreich80}
 \begin{equation}
 \left(\alpha\frac{db_{1/2}^{(m)}}{d\alpha} + 2mb_{1/2}^{(m)}\right) 
 \approx \frac{2m}{\pi}\left[ K_1\left(\frac{2}{3}\right) + 2K_0\left(\frac{2}{3}\right)\right] 
 \approx \frac{2m}{\pi} 2.51,
 \end{equation}
 where $K_0$ and $K_1$ are modified Bessel functions, so that $c_m \approx 2.55 m^2\left(1-1/m\right)$ \citep[see also][]{salmon12}. 
 
 To implement the effect of resonant torques on the disk, we use a different approach than that of \cite{charnoz10}. The disk's angular momentum $L_d$ can be written as the sum of the angular momentum of each of its cells $L_d=\sum_{j=1}^N \sigma_jS_j\sqrt{GM_PR_j}$, where $S_j$ and $R_j$ are the surface area and central position of bin number $j$. To apply the torque on the disk, we first determine the bin $j$ in which the resonance falls. For an inner resonance, the torque will cause an inward flux of mass from bin $j$ to bin $j-1$. The resulting change in the disk angular momentum $\Delta L_m$ due to the $(m:m-1)$ inner Lindblad resonance is 
 
 \begin{equation}
 \Delta L_m = L_d' - L_d = \sqrt{GM_P}\left[\left(\sigma_j'-\sigma_j\right)S_j\sqrt{R_j}+\left(\sigma_{j-1}'-\sigma_{j-1}\right)S_{j-1}\sqrt{R_{j-1}}\right],
 \end{equation}
 where primed symbols indicate resulting quantities after the torque is applied. Conservation of mass implies that $\left(\sigma_j'-\sigma_j\right)S_j = - \left(\sigma_{j-1}'-\sigma_{j-1}\right)S_{j-1}$. Finally, using $\Gamma_m =dL/dt \approx \Delta L_m / \Delta t$, the resulting surface density of bin $j$ can be written as
  
 \begin{equation}
 \sigma_j' = -\frac{\Gamma_m \Delta t}{\sqrt{GM_P}S_j\left(\sqrt{R_j}-\sqrt{R_{j-1}}\right)}+\sigma_j.\label{equ_sat_torque1}
 \end{equation}
 
 Similarly, we can write that 

 \begin{equation}
\sigma_{j-1}' = -\frac{\Gamma_m \Delta t}{\sqrt{GM_P}S_{j-1}\left(\sqrt{R_{j-1}}-\sqrt{R_j}\right)}+\sigma_{j-1}.\label{equ_sat_torque2}
\end{equation}

We are assuming that the torque from a given resonance is entirely deposited locally. In reality, the resonant torque will generate a wave that will propagate away from the perturber. However, the wave is progressively damped due to the disk's viscosity. For example, in the case of a viscous disk such as Saturn's rings, most of the torque is deposited in the region $\xi < 2$, where $\xi=\epsilon^{-1/2}\Delta R_L/R_L$ and $\Delta R_L$ is the distance to the resonance located at $R=R_L$ \citep{shu85}. The factor $\epsilon$ is a function of the local surface density and distance of the resonance. In Saturn's rings, $\epsilon$ is typically $10^{-8}$ \citep{shu84}, such that most of the torque is deposited within $\Delta R_L < 2\times10^{-4}R_L$.

To determine the evolution of the satellite's orbit due to the resonant torque, we follow the approach of \cite{papaloizou00} \citep[as in][]{salmon12}. We define an orbital migration timescale $t_{mig}=L_S/\Gamma_m$, where $L_S$ is the satellite's orbital angular momentum. This is used to calculate an acceleration on the satellite $\mathbf{a}_{mig}=\mathbf{v}/t_{mig}$, where $\mathbf{v}$ is the satellite's velocity. The disk torque on the satellite results in an additional ``kick'' that modifies the velocity following $\mathbf{v}'=\mathbf{v}+\mathbf{a}_{mig}\Delta t$.

 \subsection{Accretion outside the Roche limit}
Viscous spreading can bring disk material through the Roche limit, whose position $a_R$ is provided to the integrator as an input parameter. When this happens, we trigger the formation of a new N-body particle. The mass of the particle is set to the larger of the two following quantities: 1) the entire mass in the cell containing the Roche limit, or 2) the mass of the fragment that would form from gravitational instabilities \citep{goldreich73}
\begin{equation}
m_f=\frac{16\pi^4 \xi^3 \sigma_R^3a_R^6}{M_P^2},
\end{equation}
where $\sigma_R$ is the disk surface density at the Roche limit, and $\xi$ is a factor of order, but less than, unity. We use the largest of the two criteria to prevent the creation of an unmanageable number of small bodies. The corresponding mass is removed from the disk's cell containing the Roche limit (and sometimes inner neighboring cells when the second criterion is used), and the new body's semi-major axis is set so as to conserve the angular momentum of the system. Finally, we set the the particle's eccentricity to be the ratio of its escape velocity to the local orbital velocity \citep{lissauer93,salmon12}, and its inclination to zero (expected for formation from an equatorial disk).

We have also implemented a \textit{direct accretion} mode to replicate the ``continuous regime'' from \cite{crida12} in which a satellite orbiting close to a disk could grow from direct accretion of disk material. If this option is enabled, when disk material is being brought to the Roche limit we look for the existence of a nearby satellite. The threshold we have selected is to require the disk's edge to be within two Hill radius of the satellite, as in \cite{crida12}. If there is no such moon, then we proceed as previously explained. If there is such a qualifying satellite, then we add the disk mass directly onto it and adjust its orbital parameters to conserve angular momentum.

\subsection{Tides}\label{subsectides}
The acceleration on a satellite of mass $M_S$ due to tides raised by the satellite on a planet with mass $M_P$, radius $R_P$, spin vector $\mathbf{s_P}$ and Love number $k_{2P}$ is given by \citep{mignard80,touma94}:
\begin{equation}
\left.\frac{\partial^2\mathbf{r}}{\partial t^2}\right|_P=-\frac{3k_{2P}GMR_P^5}{r^{10}}\left(1+\frac{M_S}{M_P}\right)\Delta t_P \left[2\left(\mathbf{r}\cdot\mathbf{v}\right)\mathbf{r}+r^2\left(\mathbf{r}\times\mathbf{s_P}+\mathbf{v}\right)\right],
\end{equation}
where $\mathbf{r}=\left(x,y,z\right)$ and $\mathbf{v}=\left(v_x,v_y,v_z\right)$ are the satellite's position and velocity in a reference frame centered on the planet. The lag time $\Delta t_P$ is defined as the fixed time between the tide-raising potential and the rise of the tidal distortion. This parameter is planet-dependent and is supplied as an input to the integrator.

We evolve the spin of the planet due to the tidal torque exerted by the satellite. This torque can be written \citep{mignard80,touma94}\footnote{There is a typo in Equation (111) of \cite{touma94}: the $r_{em}^{10}$ on the denominator should be $r_{em}^8$.}
\begin{equation}
\mathbf{T}=-\frac{3k_{2P}GM_P^2R_P^5}{r^8}\Delta t_P \left[\left(\mathbf{r}\cdot\mathbf{s_P}\right)\mathbf{r}-r^2\mathbf{s_P}+\mathbf{r}\times\mathbf{v}\right].
\end{equation}
Over a time step $dt$, the tides kick the planet's spin angular momentum by $\Delta L_P=Tdt$, where $T=\left\Vert\mathbf{T}\right\Vert$. Using $\Delta L_P=\lambda_P M_P R_P^2 \Delta s_P$, where $\lambda_P$ is the planet's moment of inertia constant, the variation of the planet's spin due to tides over $dt$ is
\begin{equation}
\Delta s_P=\frac{Tdt}{\lambda_P M_P R_P^2}.
\end{equation}
The parameter $\lambda_P$ is supplied to the integrator as an input parameter. The planet's radius is usually set constant, but it can also be progressively adjusted to, e.g., account for the planet contracting \citep[as in ][]{salmon17}, over a timescale supplied in input. In such cases, the planet's spin is also adjusted to conserve the rotational angular momentum.

Similarly, the acceleration on the satellite due to tides raised by the planet on the satellite can be written as \citep{mignard80,touma94}:
\begin{equation}
\left.\frac{\partial^2\mathbf{r}}{\partial t^2}\right|_S=-\frac{3GM_Pk_{2P}R_P^5}{r^{10}}\left(1+\frac{M_S}{M_P}\right)\mathcal{A}\Delta t_P \left[2\left(\mathbf{r}\cdot\mathbf{v}\right)\mathbf{r}+r^2\left(\mathbf{r}\times\mathbf{s_S}+\mathbf{v}\right)\right],
\end{equation}
where $\mathbf{s_S}$ is the satellite's spin vector. The factor $\mathcal{A}$ reflects the strength of satellite versus planetary tides, with
\begin{equation}
\mathcal{A}=\left(\frac{M_S}{M_P}\right)^{-2} \left(\frac{R_S}{R_P}\right)^5 \left(\frac{k_{2S}}{k_{2P}}\right) \left(\frac{\Delta t_S}{\Delta t}\right),
\end{equation}
where $k_{2S}$, $\Delta t_S$ and $R_S$ are the satellite's Love number, tidal time lag and physical radius. When implementing the latter equation in our integrator, we assume that the satellite is rotating synchronously, so that $s_S \approx n=\sqrt{GM_P/a^3}$, where $a$ is the satellite's semi-major axis. This is a reasonable assumption, given that despinning timescales are generally short compared to tidal evolution timescales \citep[e.g.][]{salmon17}

\subsection{Control of the time step}
The overall integration timestep $dt$ is set initially by SyMBA’s requirement to be no larger than 1/20th of the innermost orbital period to be considered. The disk's evolution is calculated for half a time step $dt/2$ before and after the \textit{N}-body integration is performed. However, this timestep can be too big for the hydrocode, and can lead to instabilities and negative surface densities. We thus allow for subdividing the timestep for the evolution of the disk calculation. The Diffusive Stability Constraint (DSC) requires the timestep to be no larger than 
\begin{equation}
dt_{DSC}=\frac{1}{2}\frac{\Delta R^2}{\nu_{Max}},
\end{equation}
where $\Delta R$ is the grid's radial resolution, and $\nu_{Max}$ is the maximum value of the viscosity throughout the entire disk \citep{Johnston04}. Through our testing we found that limiting the timestep of the hydrocode to $dt_{hydro}=0.1*dt_{DSC}$ ensures stability of the integrator. In practice, at each iteration we calculate $dt_{hydro}$. If it is larger than $dt/2$, we set $dt_{hydro}=dt/2$. Otherwise, we perform an integration using $dt_{hydro}$ and repeat the operation until the disk's evolution has been computed for a full $dt/2$.

\section{Tests}
\subsection{Viscous spreading}
\subsubsection{Constant viscosity}
We perform a first series of tests in which we only consider the viscous spreading of a disk, in the absence of any orbiting bodies. In order to assess the accuracy of the code, we perform a first test using a constant and uniform viscosity model and compare the disk's evolution to the analytical result from \cite{pringle81}. The latter is a solution for an initial Dirac distribution centered at $r=r_0$, which we cannot exactly replicate due to the finite resolution of the grid. We approximate this distribution by setting the disk's surface density to 0 except in the bin $i$ containing $r=r_0$, where we set $\sigma=M_d/S_i$. We set the primary to be the Earth, and the disk mass is equal to one Lunar mass $\left(M_L\right)$. We use a radial grid extending from 0 to $\SI{2.4}{R_{\oplus}}$ with 240 equal-sized bins. The boundary conditions are set to $F_1 = F_2$ and $F_{N+1} = F_N$, which allows material to flow freely through the grid's edges. We set the Roche limit to an artificially large value in order to de-activate the creation of moonlets.

\begin{figure}[h!]
	\begin{center}
		\includegraphics[width=8.5cm]{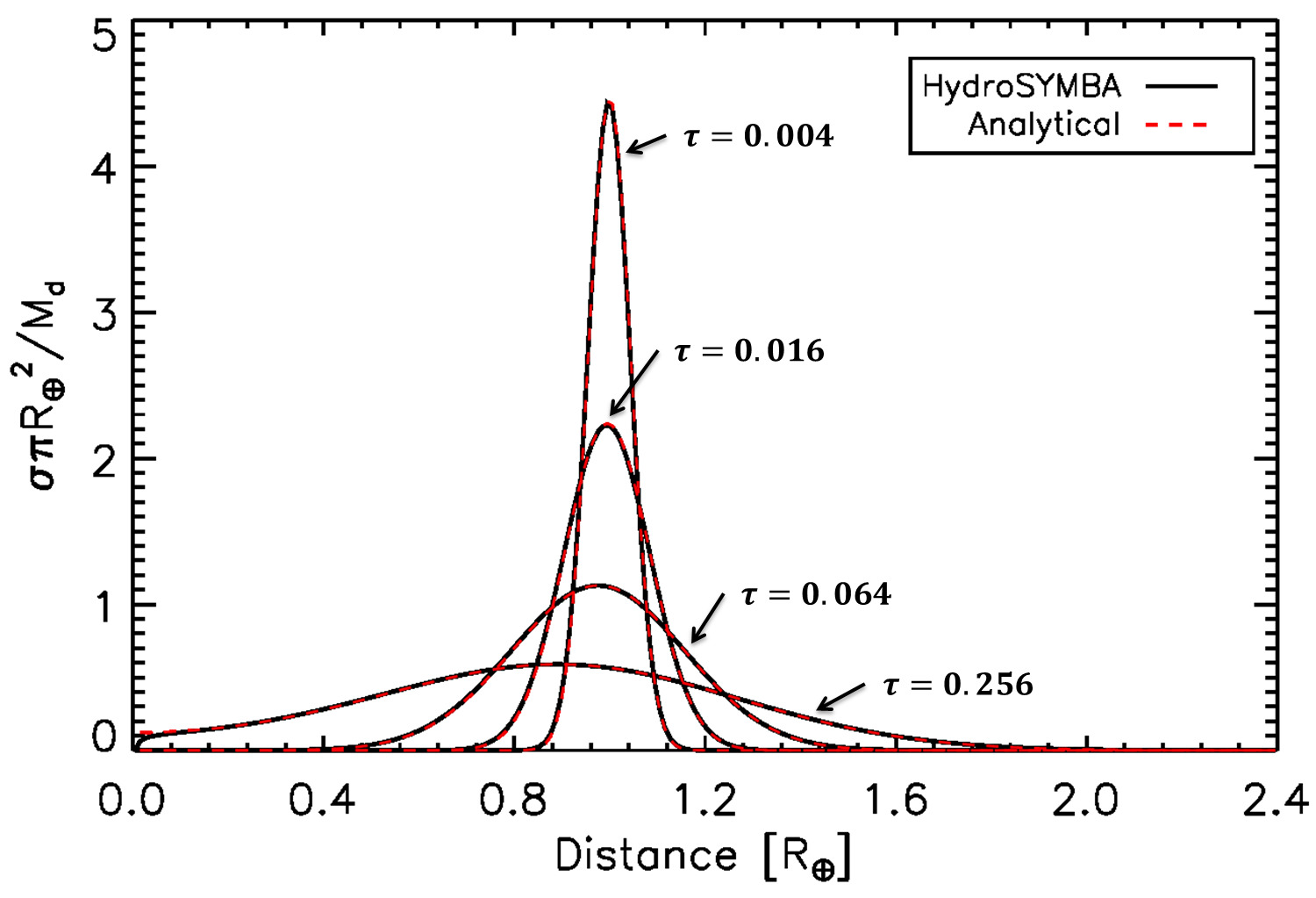}
		\caption{Viscous spreading of a disk with constant viscosity. The black curves show the disk's surface density at different times using our new integrator. The red dashed lines show the analytical solution from Equation \ref{equ_pringle}. \label{fig_pringle_test}}
	\end{center}
\end{figure}

The surface density as a function of time and distance is given by \cite{pringle81}:

\begin{equation}
\sigma \left(x,\tau\right)=\frac{M_d}{\pi r_0^2}\tau^{-1}x^{-1/4}\exp^{-\left(1+x^2\right)/\tau}I_{1/4}\left(2x/\tau\right), \label{equ_pringle}
\end{equation}
where $x=r/r0$, $\tau=12\nu t/r_0^2$, and $I_{1/4}$ is a modified Bessel function of the first kind. Figure \ref{fig_pringle_test} shows the surface density profile calculated by the integrator in black, and the analytical solution from Equation \ref{equ_pringle} in red dashes, at four times of evolution. Our results are in excellent agreement with the analytical solution.

\subsubsection{Variable viscosity}
\begin{figure}[h!]
 \begin{center}
 	\includegraphics[width=17cm]{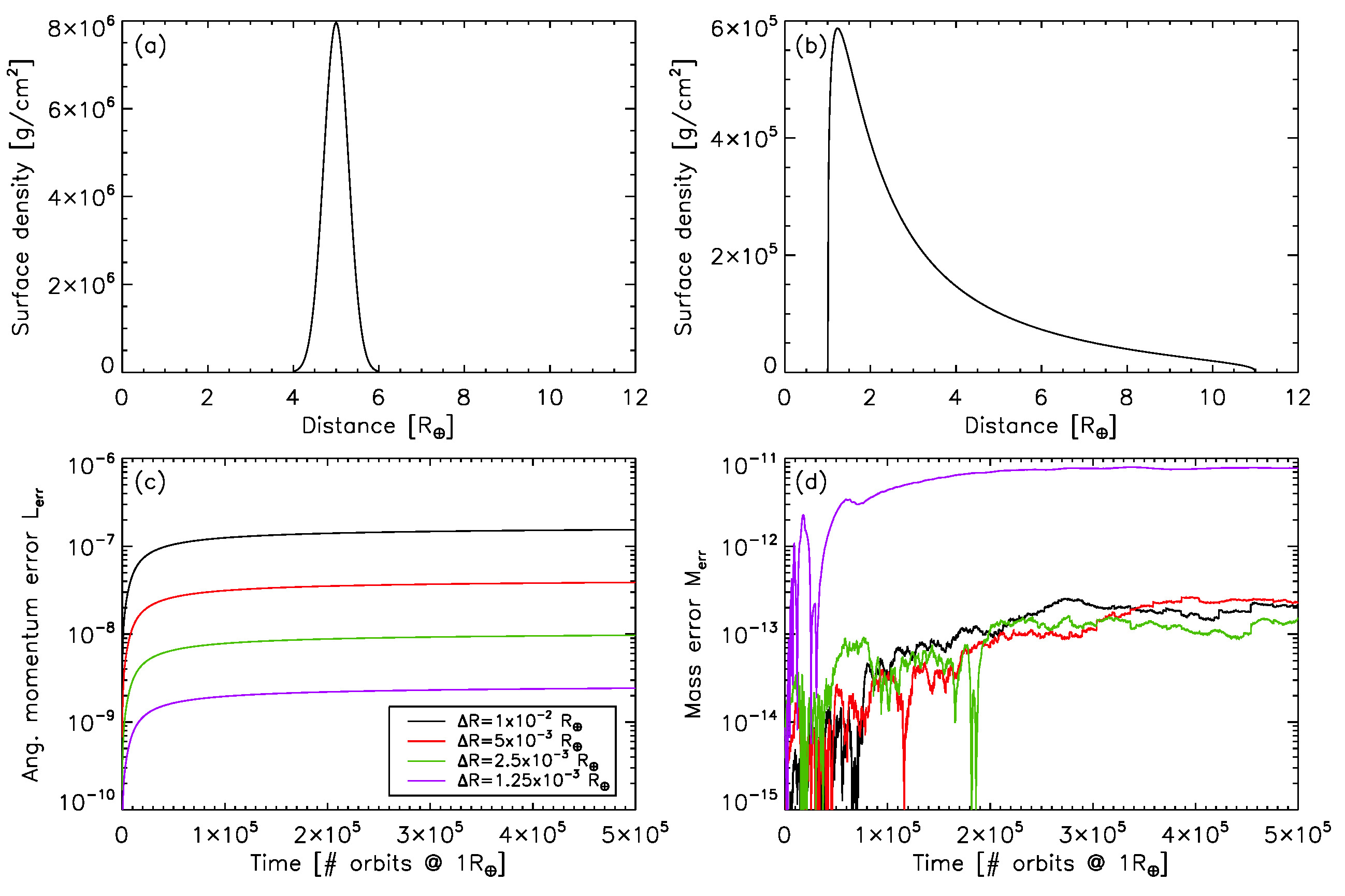}
 	\caption{(a) Initial surface density profile. (b) Disk's surface density at $t=5\times10^5 T_K$ for the 1000 bins case. (c) Angular momentum error $L_{err}$ as a function of time, for a grid with 1000 bins (black line), 2000 bins (red line), 4000 bins (green line), and 8000 bins (purple line). (d) Mass error $M_{err}$ as a function of time, with colors identical to panel (c). The larger error for the 8000 bins case is the result of rounding errors due to the smallness of the DSC-constrained timestep.\label{fig_viscous_spreading}}
 \end{center}
 \end{figure}

We use a similar setup to the previous section, but this time start with a Gaussian surface density distribution between 4 and $6R_{\oplus}$ (Figure \ref{fig_viscous_spreading}, panel (a)). We use a radial grid extending from 1 to $11R_{\oplus}$. We perform 4 simulations with 1000, 2000, 4000 and 8000 grid cells, which correspond to radial resolutions $\Delta R$ of $10^{-2}$, $5\times 10^{-3}$, $2.5\times10^{-3}$, and $\SI{1.25e-3}{R_\oplus}$, respectively. We use a gravitational instability viscosity model \citep{ward78}:

\begin{equation}
\nu_{WC} = \frac{\pi^2G^2\sigma^2}{\Omega^3}
\end{equation}

We integrate the system for $5\times 10^5 T_K$, where $T_K$ is the orbital period at $1R_{\oplus}$. This integration time corresponds to about $3\times 10^6$, $1.5\times 10^7$, $6\times 10^7$ and $2.4\times 10^8$ times the DSC timestep at t=0 for the 4 grid resolutions. Figure \ref{fig_viscous_spreading} shows the disk's surface density at $t=0$ in panel (a), and at $t=5\times10^5T_K$ in panel (b), for the N=1000 case. Panels (c) and (d) show the angular momentum error $L_{err}$ and mass error $M_{err}$, which we define as:
\begin{align}
 L_{err}=\frac{L_d-L_{d,0}+L_{pl}+L_{out}}{L_{d,0}}\\
 M_{err}=\frac{M_d-M_{d,0}+M_{pl}+M_{out}}{M_{d,0}}
 \end{align}
 where $L_{d,0}$ and $M_{d,0}$ are the disk's initial angular momentum and mass, and $L_{pl}$ and $L_{out}$ are the angular momentum of the mass falling on the planet $M_{pl}$ and being removed through the outer edge $M_{out}$, respectively. The error in the angular momentum of the disk depends strongly on the resolution, but is small in all cases. At $t=5\times10^5 T_K$, the angular momentum error is $1.55\times10^{-7}$, $3.88\times10^{-8}$, $9.72\times10^{-9}$ and $2.43\times10^{-9}$ for the grids with 1000, 2000, 4000 and 8000 bins, respectively. The error decreases as $N^2$, i.e. increasing the resolution by a factor of 2 decreases the angular momentum error by a factor of 4.
 
 We find that the integrator conserves mass to machine precision with all resolutions, except the $N=8000$ cells case which is significantly worse than the other three coarser resolutions. We believe this is a consequence of rounding errors due to the integration timestep becoming too small in order to satisfy the DSC (which is a function of $\Delta R^2$). Our algorithm is currently written in double precision. While implementing quadruple precision might allow for finer grid resolutions (which might be desirable when satellites are included in order to accurately track deposition of the resonant torque in the disk), we believe that the accuracy of the integrator with slightly coarser resolutions is satisfactory for the most desired purposes.
 
\subsection{Disk and satellite}
\begin{figure}[h!]
	\begin{center}
		\includegraphics[width=17cm]{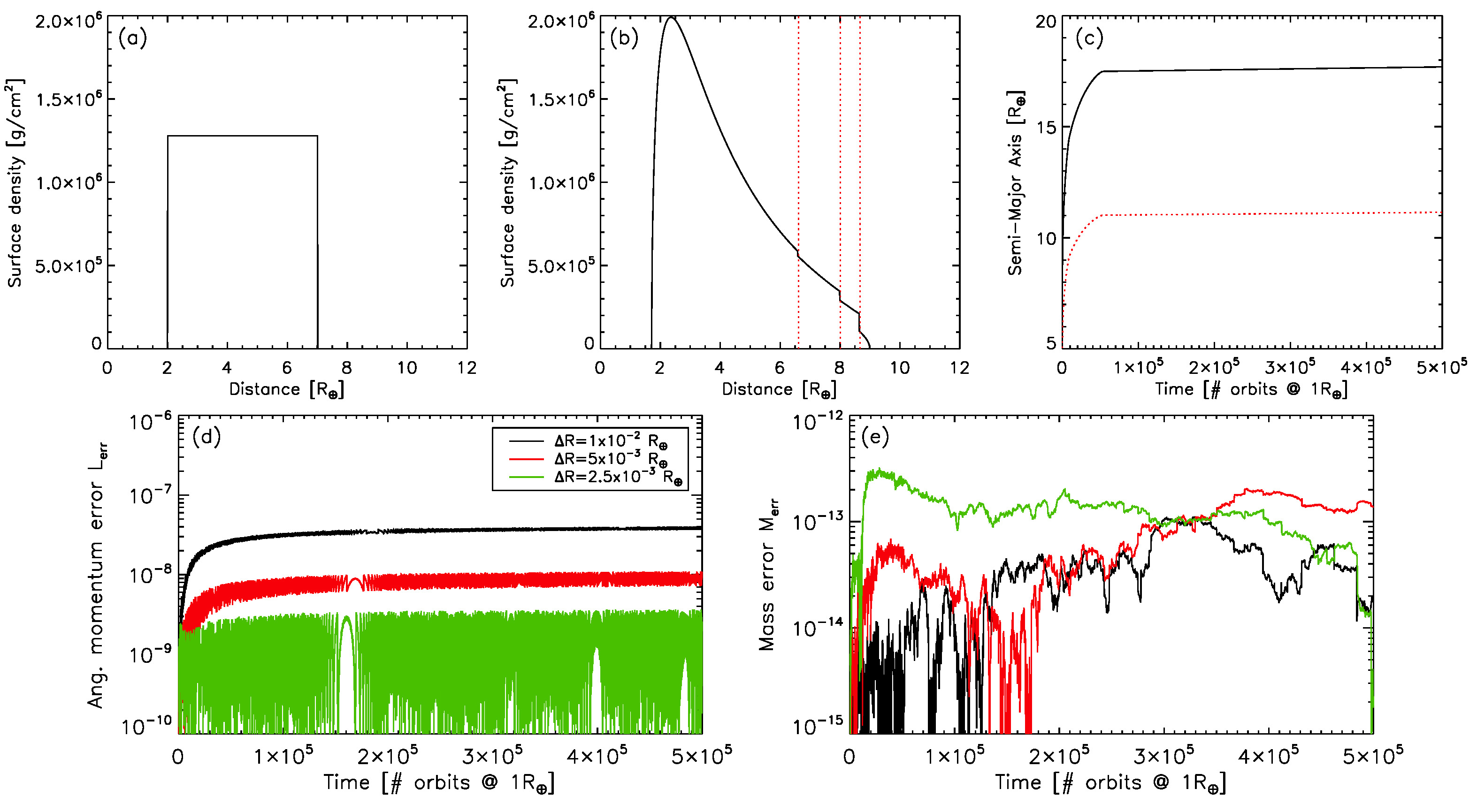}
		\caption{(a) Initial surface density profile. (b) Disk's surface density at $t=1000 T_K$ for the 1000 bins case. The red lines show the positions of the satellite's mean-motion resonances. (c) Evolution of the satellite's semi-major axis (black line), and position of its 2:1 mean-motion resonance (red line) The satellite's outward migration slows down once its 2:1 resonance lies at the edge of the disk at $11 R_\oplus$, after which its outward motion is controlled by the primary's tidal torque. (d) Angular momentum error $L_{err}$ as a function of time, for a grid with 1000 bins (black line), 2000 bins (red line), and 4000 bins (green line). (e) Mass error $M_{err}$ as a function of time, with colors identical to panel (d).\label{fig_angmom_error_1bod}}
	\end{center}
\end{figure}

In this second test we consider a slab of material extending from 2 to $7R_\oplus$ around an Earth-mass planet, with an initial mass of $1M_L$. We consider radial grids identical to the previous case, but do not consider the $N=8000$ bins case. Our boundary conditions are again set to let material flow freely through the inner and outer edges, and we again use the viscosity model from \cite{ward78}. The Roche limit is again set to an artificially large value in order to de-activate the creation of moonlets. We place a $0.1 M_L$ satellite at an initial semi-major axis of $8R_\oplus$. We turn on the tidal dissipation in the planet, using a spin period of 5 hours (typical of a post-giant impact Earth), and a tidal lag $\Delta t_P = \SI{6.9e6}{\second}$ (about 1000 times the current tidal lag of the Earth), and integrate the system for $5\times10^5T_K$. Figure \ref{fig_angmom_error_1bod} shows the disk at t=0 and $1000T_K$ in panels a) and b). In the latter, note how the disk's surface density is sculpted by the satellite's resonances, plotted as the vertical red lines. Panel c) shows the evolution of the satellite's semi-major axis and position of its 2:1 resonance. The satellite moves outward rapidly until it reaches $\sim 17.46R_\oplus$, at which point its 2:1 resonance lies just outside of the disk's outer edge at  $11R_\oplus$. It then continues to move outward slowly due to the primary's tidal torque. The final semi-major axis of the satellite is $\SI{17.691}{R_\oplus }$, $\SI{17.698}{R_\oplus }$ and $\SI{17.702}{R_\oplus }$ with N=1000, 2000 and 4000 bins.

Panels d) and e) show conservation of the angular momentum and mass of the system (disk + satellite). In this case, we define the angular momentum error as $L_{err}=\left( L_{tot}(t)-L_{tot,0}\right)/L_{tot,0}$, with $L_{tot}=L_d+L_{pl}+L_{out}+L_P+L_{orb}$, where ${L_P}$ is the spin angular momentum of the planet and $L_{orb}$ is the orbital angular momentum of the satellite. Our integrator again conserves mass down to machine precision with all 3 resolutions. Conservation of angular momentum is similar to the purely viscous case. For comparison, the widely used code Fargo conserves angular momentum in the heliocentric frame exactly by construction, but the total angular momentum in the barycentric frame only to $\sim 10^{-3}$ precision over $\sim 2500T_K$ in the case of a Jupiter mass planet embedded in a protoplanetary disk \citep{crida07}.

\subsection{Lunar accretion}
\begin{figure}[h!]
	\begin{center}
		\includegraphics[width=17cm]{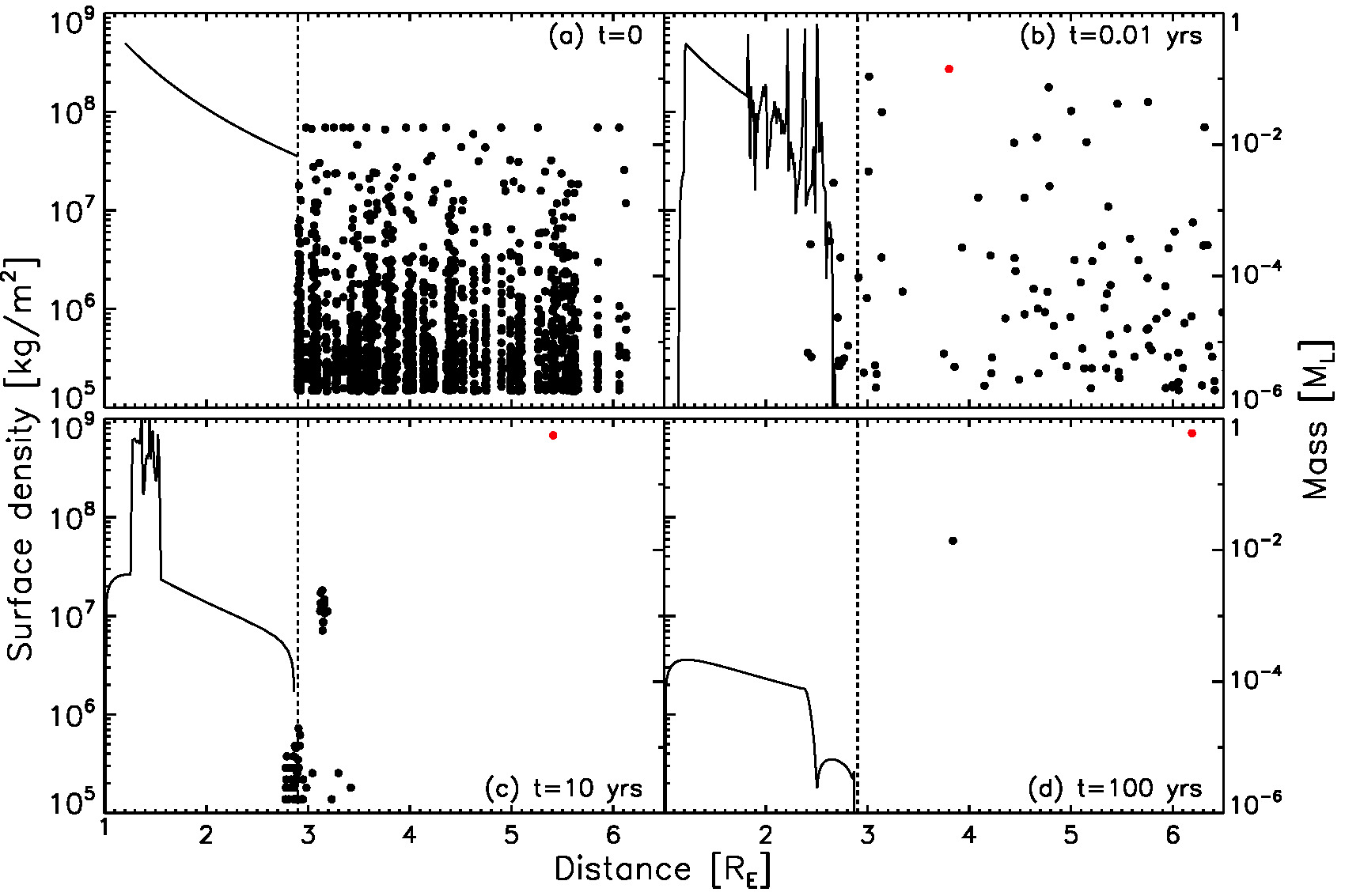}
		\caption{Accretion from a protolunar disk. On all panels, the left axis shows the surface density of the Roche-interior disk, while the right axis shows the mass of individual particles. The material in the outer disk rapidly accretes into a few large objects. This initial phase results in significant confinement of the inner disk well interior to the Roche limit (vertical dashed line). Eventually, the disk viscously spreads back out and can produce new moonlets that continue the growth of the Moon (red dot in panels b-d). The sharp surface density transitions in panel (c) are due to viscosity transitions. As the proto-Moon scatters inner moonlets or captures them into mean motion resonances, it continues to gain angular momentum and its semi-major axis expands beyond what could be achieved solely by interacting with the Roche-interior disk. \label{fig_full_accretion}}
	\end{center}
\end{figure}

We perform a final test in which we model the evolution of a protolunar disk, and the accumulation of the Moon from disk material. The Roche-interior disk contains $\SI{1.5}{M_L}$ between 1.2 and $\SI{2.9}{R_\oplus}$, with a power-law surface density profile $\sigma(r) \propto r^{-3}$. The radial grid contains 200 bins extending from 1 to $\SI{3}{R_\oplus}$, corresponding to a resolution $\Delta R=\SI{e-2}{R_\oplus}$. The outer disk contains 2000 bodies following a cumulative size-frequency distribution $N(>R)\propto R^{-1.5}$, totaling $\SI{0.5}{M_L}$ and extending from 3 to $\SI{6}{R_\oplus}$. Figure \ref{fig_full_accretion}a shows the initial disk. We use a viscosity model appropriate for a two-phase protolunar disk \citep{thompson88,salmon12} $\nu=\min\left(\nu_{WC},\nu_{TS}\right)$, with

\begin{equation}
\nu_{TS}=\frac{\sigma_{SB}T_P^4}{\sigma \Omega^2},
\end{equation}
where $\sigma_{SB}$ is the Stefan-Boltzmann constant and $T_P$ is the vapor disk's photospheric temperature. We use $T_P=\SI{2000}{K}$, appropriate for a vertically unstratified silicate vapor disk \citep{thompson88}. Finally, we do not include tidal interactions.

As found in previous works \citep[e.g.][]{salmon12}, the accretion of the material in the outer disk occurs in less than a year (Figure \ref{fig_full_accretion}b), while delivery of inner-disk material occurs over decades. We find that as the growing Moon scatters inner moonlets, it gains significant angular momentum, which expands its semi-major axis well beyond what could be achieved solely via direct disk-satellite interactions ($\sim\SI{4.6}{r_\oplus}$). Capture into mean-motion resonances is also efficient, with the two bodies in Figure \ref{fig_full_accretion}d in 2:1 resonance.

\section{Summary}
We have developed a new numerical integrator that allows one to model the evolution of a continuous disk and individual bodies, within a single, self-consistent environment. The evolution of the disk is computed using a second-order Runge-Kutta integration scheme, while the orbital evolution of the particles is handled by the symplectic N-body integrator SyMBA. The disk evolves due to its viscosity (with various viscosity models available), and with satellite interactions through first-order Lindblad resonances. The evolution of the satellites includes particle-particle interactions (including collision and mergers), back reaction from resonant interactions with the disk, and tidal dissipation in the primary and/or the satellite itself. 

The current integrator represents a significant step forward in the modeling on circumplanetary disks and satellites. It is suitable to study a variety of systems, e.g.: a well-mixed non-stratified protolunar disk; the rings and satellites of Saturn; the formation of Phobos and Deimos from a circum-Mars disk; the formation of some of Uranus's satellites \citep{morbidelli12b}; the formation of the moons of exoplanets \citep{nakajima14}. We have also identified some additional features that would be useful to incorporate

\begin{itemize}
	\item \textbf{Tidal disruption} In the current integrator, an object scattered inside the Roche limit remains intact. However, it may instead be tidally disrupted. \cite{sridhar92} indicate that an object of density $\rho$ will be entirely disrupted in a single pass if its pericenter $q<1.05\left(M_P/\rho\right)^{(1/3)}$. This can be a significant process as shown in simulations of lunar accretion in \cite{salmon12}. Alternatively, a body would be absorbed by the inner disk if as it passed through the disk it encountered a mass comparable to or greater than its own.
	\item \textbf{Layered disk} Models of lunar accretion consider disks in which the vapor and melt phases are well-mixed and co-evolve \citep{salmon12,salmon14}. However, recent works \citep{ward12} argue that the condensed phase of the protolunar disk rapidly settles in the midplane, resulting in a disk of magma surrounded by a vapor atmosphere. The latter can be represented by two additional disks, above and below the condensed disk. The evolution of each phase needs then to be modeled separately as they each will respond to different physical processes, or to similar processes but with different characteristic parameters.
	\item \textbf{Irregular grid} It would be valuable to implement grids with variable resolution, as it would allow one to model more precisely certain regions of the disk (e.g. the outer region where most resonant interactions occur) at a reduced computational cost.
\end{itemize}

\section*{Acknowledgements}
This work has been funded by Southwest Research Institute Internal Research program, and NASA's Solar System Exploration Research Virtual Institute grant NNA14AB03A. We thank an anonymous referee for their careful reading of the paper and their valuable comments. Southwest Research Institute's Internal Research program provided primary funding for the code's development. While the code ir proprietary to SwRI, we are happy to make it available to other scientists in the context of mutually collaborative projects. We invite interested readers to contact us directly for such requests.

\bibliography{references}

\begin{thebibliography}{}
\expandafter\ifx\csname natexlab\endcsname\relax\def\natexlab#1{#1}\fi

\bibitem[{{Cameron} \& {Ward}(1976)}]{cameron76}
{Cameron}, A.~G.~W., \& {Ward}, W.~R. 1976, in Lunar and Planetary
  Inst.~Technical Report, Vol.~7, Lunar and Planetary Science Conference

\bibitem[{{Canup} \& {Salmon}(2018)}]{canup18}
{Canup}, R., \& {Salmon}, J. 2018, Science Advances, 4, eaar6887

\bibitem[{{Canup}(2010)}]{canup10}
{Canup}, R.~M. 2010, \nat, 468, 943

\bibitem[{{Canup} \& {Esposito}(1995)}]{canup95}
{Canup}, R.~M., \& {Esposito}, L.~W. 1995, \icarus, 113, 331

\bibitem[{{Canup} \& {Ward}(2006)}]{canup06}
{Canup}, R.~M., \& {Ward}, W.~R. 2006, \nat, 441, 834

\bibitem[{{Charnoz} {et~al.}(2010){Charnoz}, {Salmon}, \& {Crida}}]{charnoz10}
{Charnoz}, S., {Salmon}, J., \& {Crida}, A. 2010, \nat, 465, 752

\bibitem[{{Charnoz} {et~al.}(2011){Charnoz}, {Crida}, {Castillo-Rogez},
  {Lainey}, {Dones}, {Karatekin}, {Tobie}, {Mathis}, {Le Poncin-Lafitte}, \&
  {Salmon}}]{charnoz11}
{Charnoz}, S., {Crida}, A., {Castillo-Rogez}, J.~C., {et~al.} 2011, \icarus,
  216, 535

\bibitem[{{Crida} \& {Charnoz}(2012)}]{crida12}
{Crida}, A., \& {Charnoz}, S. 2012, Science, 338, 1196

\bibitem[{{Crida} {et~al.}(2007){Crida}, {Morbidelli}, \& {Masset}}]{crida07}
{Crida}, A., {Morbidelli}, A., \& {Masset}, F. 2007, \aap, 461, 1173

\bibitem[{{Daisaka} {et~al.}(2001){Daisaka}, {Tanaka}, \& {Ida}}]{daisaka01}
{Daisaka}, H., {Tanaka}, H., \& {Ida}, S. 2001, \icarus, 154, 296

\bibitem[{{Duncan} {et~al.}(1988){Duncan}, {Quinn}, \& {Tremaine}}]{duncan88}
{Duncan}, M., {Quinn}, T., \& {Tremaine}, S. 1988, \apjl, 328, L69

\bibitem[{{Duncan} {et~al.}(1998){Duncan}, {Levison}, \& {Lee}}]{duncan98}
{Duncan}, M.~J., {Levison}, H.~F., \& {Lee}, M.~H. 1998, \aj, 116, 2067

\bibitem[{{Goldreich} \& {Tremaine}(1980)}]{goldreich80}
{Goldreich}, P., \& {Tremaine}, S. 1980, \apj, 241, 425

\bibitem[{{Goldreich} \& {Tremaine}(1978)}]{goldreich78}
{Goldreich}, P., \& {Tremaine}, S.~D. 1978, \icarus, 34, 227

\bibitem[{{Goldreich} \& {Ward}(1973)}]{goldreich73}
{Goldreich}, P., \& {Ward}, W.~R. 1973, \apj, 183, 1051

\bibitem[{{Hesselbrock} \& {Minton}(2017)}]{Hesselbrock17}
{Hesselbrock}, A.~J., \& {Minton}, D.~A. 2017, Nature Geoscience, 10, 266

\bibitem[{{Hyodo} {et~al.}(2017){Hyodo}, {Rosenblatt}, {Genda}, \&
  {Charnoz}}]{hyodo17}
{Hyodo}, R., {Rosenblatt}, P., {Genda}, H., \& {Charnoz}, S. 2017, \apj, 851,
  122

\bibitem[{Johnston \& Liu(2004)}]{Johnston04}
Johnston, H., \& Liu, J.-G. 2004, J. Comput. Phys., 199, 221

\bibitem[{{Lissauer} \& {Stewart}(1993)}]{lissauer93}
{Lissauer}, J.~J., \& {Stewart}, G.~R. 1993, in Astronomical Society of the
  Pacific Conference Series, Vol.~36, Planets Around Pulsars, ed.
  {J.~A.~Phillips, S.~E.~Thorsett, \& S.~R.~Kulkarni}, 217--233

\bibitem[{{Lynden-Bell} \& {Pringle}(1974)}]{lynden74}
{Lynden-Bell}, D., \& {Pringle}, J.~E. 1974, \mnras, 168, 603

\bibitem[{{McNeil} {et~al.}(2005){McNeil}, {Duncan}, \& {Levison}}]{mcneil05}
{McNeil}, D., {Duncan}, M., \& {Levison}, H.~F. 2005, \aj, 130, 2884

\bibitem[{{Mignard}(1980)}]{mignard80}
{Mignard}, F. 1980, Moon and Planets, 23, 185

\bibitem[{{Morbidelli} \& {Nesvorny}(2012)}]{morbidelli12a}
{Morbidelli}, A., \& {Nesvorny}, D. 2012, \aap, 546, A18

\bibitem[{{Morbidelli} {et~al.}(2012){Morbidelli}, {Tsiganis}, {Batygin},
  {Crida}, \& {Gomes}}]{morbidelli12b}
{Morbidelli}, A., {Tsiganis}, K., {Batygin}, K., {Crida}, A., \& {Gomes}, R.
  2012, \icarus, 219, 737

\bibitem[{{Nakajima} {et~al.}(2014){Nakajima}, {Genda}, {Asphaug}, \&
  {Ida}}]{nakajima14}
{Nakajima}, M., {Genda}, H., {Asphaug}, E., \& {Ida}, S. 2014, in AAS/Division
  for Planetary Sciences Meeting Abstracts, Vol.~46, AAS/Division for Planetary
  Sciences Meeting Abstracts \#46, 201.03

\bibitem[{{Papaloizou} \& {Larwood}(2000)}]{papaloizou00}
{Papaloizou}, J.~C.~B., \& {Larwood}, J.~D. 2000, \mnras, 315, 823

\bibitem[{{Pringle}(1981)}]{pringle81}
{Pringle}, J.~E. 1981, \araa, 19, 137

\bibitem[{{Rosenblatt} {et~al.}(2016){Rosenblatt}, {Charnoz}, {Dunseath},
  {Terao-Dunseath}, {Trinh}, {Hyodo}, {Genda}, \& {Toupin}}]{rosenblatt16}
{Rosenblatt}, P., {Charnoz}, S., {Dunseath}, K.~M., {et~al.} 2016, Nature
  Geoscience, 9, 581

\bibitem[{{Salmon} \& {Canup}(2012)}]{salmon12}
{Salmon}, J., \& {Canup}, R.~M. 2012, \apj, 760, 83

\bibitem[{{Salmon} \& {Canup}(2014)}]{salmon14}
---. 2014, Royal Society of London Philosophical Transactions Series A, 372,
  30256

\bibitem[{{Salmon} \& {Canup}(2017)}]{salmon17}
---. 2017, \apj, 836, 109

\bibitem[{{Salmon} {et~al.}(2010){Salmon}, {Charnoz}, {Crida}, \&
  {Brahic}}]{salmon10}
{Salmon}, J., {Charnoz}, S., {Crida}, A., \& {Brahic}, A. 2010, \icarus, 209,
  771

\bibitem[{{Salo}(1992)}]{salo92}
{Salo}, H. 1992, Nature, 359, 619

\bibitem[{{Salo}(1995)}]{salo95}
---. 1995, \icarus, 117, 287

\bibitem[{{Shakura} \& {Sunyaev}(1973)}]{shakura73}
{Shakura}, N.~I., \& {Sunyaev}, R.~A. 1973, \aap, 24, 337

\bibitem[{{Shu}(1984)}]{shu84}
{Shu}, F.~H. 1984, in IAU Colloq. 75: Planetary Rings, ed. R.~{Greenberg} \&
  A.~{Brahic}, 513--561

\bibitem[{{Shu} {et~al.}(1985){Shu}, {Dones}, {Lissauer}, {Yuan}, \&
  {Cuzzi}}]{shu85}
{Shu}, F.~H., {Dones}, L., {Lissauer}, J.~J., {Yuan}, C., \& {Cuzzi}, J.~N.
  1985, \apj, 299, 542

\bibitem[{Spahn \& Schmidt(2006)}]{spahn06}
Spahn, F., \& Schmidt, J. 2006, GAMM-Mitteilungen, 29, 118

\bibitem[{{Sridhar} \& {Tremaine}(1992)}]{sridhar92}
{Sridhar}, S., \& {Tremaine}, S. 1992, \icarus, 95, 86

\bibitem[{{Thompson} \& {Stevenson}(1988)}]{thompson88}
{Thompson}, C., \& {Stevenson}, D.~J. 1988, \apj, 333, 452

\bibitem[{{Touma} \& {Wisdom}(1994)}]{touma94}
{Touma}, J., \& {Wisdom}, J. 1994, \aj, 108, 1943

\bibitem[{{van Lieshout} {et~al.}(2018){van Lieshout}, {Kral}, {Charnoz},
  {Wyatt}, \& {Shannon}}]{vanlieshout18}
{van Lieshout}, R., {Kral}, Q., {Charnoz}, S., {Wyatt}, M.~C., \& {Shannon}, A.
  2018, \mnras, 480, 2784

\bibitem[{{Ward}(2012)}]{ward12}
{Ward}, W.~R. 2012, \apj, 744, 140

\bibitem[{{Ward} \& {Cameron}(1978)}]{ward78}
{Ward}, W.~R., \& {Cameron}, A.~G.~W. 1978, in Lunar and Planetary
  Inst.~Technical Report, Vol.~9, Lunar and Planetary Science Conference,
  1205--1207

\bibitem[{{Wisdom} \& {Tremaine}(1988)}]{wisdom88}
{Wisdom}, J., \& {Tremaine}, S. 1988, \aj, 95, 925

\end{thebibliography}

\end{document}